\documentclass[aps,prl,twocolumn,superscriptaddress]{revtex4-1}

\usepackage{amsmath}
\usepackage{amsfonts}
\usepackage{amssymb}
\usepackage{amsthm}
\usepackage{mathtools}

\usepackage{epsf}
\usepackage{graphicx}
\usepackage{epstopdf}

\usepackage{color}

\begin{document}

\title{Isolation of proximity-induced triplet pairing channel in a superconductor/ferromagnet spin valve}

\author{P.~V.~Leksin}
\affiliation{Leibniz Institute for Solid State and Materials
Research IFW Dresden, D-01171 Dresden, Germany}
\affiliation{Zavoisky Physical-Technical Institute, Russian
Academy of Sciences, 420029 Kazan, Russia}

\author{N.~N.~Garif'yanov}
\affiliation{Zavoisky Physical-Technical Institute, Russian
Academy of Sciences, 420029 Kazan, Russia}

\author{A.~A.~Kamashev}
\affiliation{Zavoisky Physical-Technical Institute, Russian
Academy of Sciences, 420029 Kazan, Russia}

\author{A.~A.~Validov}
\affiliation{Zavoisky Physical-Technical Institute, Russian
Academy of Sciences, 420029 Kazan, Russia}

\author{Ya.~V.~Fominov}
\affiliation{L.~D.\ Landau Institute for Theoretical Physics, Russian Academy of Sciences, 142432 Chernogolovka, Russia}
\affiliation{Moscow Institute of Physics and Technology, 141700
Dolgoprudny, Russia}

\author{J.~Schumann}
\affiliation{Leibniz Institute for Solid State and Materials
Research IFW Dresden, D-01171 Dresden, Germany}

\author{V.~Kataev}
\affiliation{Leibniz Institute for Solid State and Materials
Research IFW Dresden, D-01171 Dresden, Germany}

\author{J.~Thomas}
\affiliation{Leibniz Institute for Solid State and Materials Research IFW Dresden, D-01171 Dresden, Germany}

\author{B.~B\"{u}chner}
\affiliation{Leibniz Institute for Solid State and Materials
Research IFW Dresden, D-01171 Dresden, Germany}
\affiliation{Institute for Solid State Physics, Technical University
Dresden, D-01062 Dresden, Germany}

\author{I.~A.~Garifullin}
\affiliation{Zavoisky Physical-Technical Institute, Russian
Academy of Sciences, 420029 Kazan, Russia}

\date{\today}

\begin{abstract}
We have studied the proximity-induced superconducting triplet
pairing in CoO$_x$/Py1/Cu/Py2/Cu/Pb spin-valve structure (where Py
= Ni$_{0.81}$Fe$_{0.19}$). By optimizing the parameters of this
structure  we found a triplet-channel assisted full switching
between the normal and superconducting states. To observe an
``isolated'' triplet spin-valve effect we exploited the
oscillatory feature of the magnitude of the ordinary spin-valve
effect $\Delta T_c$ in the dependence of the Py2-layer thickness
$d_{Py2}$. We determined the value of $d_{Py2}$ at which $\Delta
T_c$ caused by the ordinary spin-valve effect (the difference in
the superconducting transition temperature $T_c$ between the
antiparallel and parallel mutual orientation of magnetizations of
the Py1 and Py2 layers) is suppressed. For such a sample a
``pure'' triplet spin-valve effect which causes the minimum in
$T_c$ at the orthogonal configuration of magnetizations has been
observed.

\end{abstract}

\pacs{74.45+c, 74.25.Nf, 74.78.Fk}

\keywords{superconductor,ferromagnet,proximity effect}

\maketitle

The superconducting  spin-valve effect consists of different
degree of suppression of superconductivity in the F1/F2/S or F1/S/F2
thin film multilayer constructions at parallel (P) and antiparallel (AP)  mutual
orientation of magnetizations of the F1 and F2 ferromagnetic layers. The
superconducting spin valves based on the superconductor/ferromagnet
(S/F) proximity effect offer a playground to explore
fundamental aspects of interplay between superconductivity and
magnetism and also promise applications as passive devices of the
superconducting spintronics. The latter construction should be
operational upon application of a small external magnetic field.
Many experimental works were performed  to confirm this effect for
the S/F systems with a good contact between metallic F and S layers
made of ordinary metals and standard ferromagnets (see, e.g., recent
reviews \cite{eschrig,blamire,Linder2015} and references therein).
In spite of different values of the magnitude of the spin-valve
effect $\Delta T_c=T_c^\mathrm{AP}-T_c^\mathrm{P}$ ($\Delta T_c=10$ mK in
Ref.~\cite{Zdravkov2013}, $\Delta T_c=20$ mK in Ref.~\cite{Jara2014} and
$\Delta T_c=120$ mK in Ref.~\cite{Wang2014}), the full switching between
the superconducting and normal states has been realized only in
a few cases \cite{leksin1,Li} because
$\Delta T_c$ was usually smaller than the width of the superconducting
transition $\delta T_c$.

Very recently, Singh \textit{et al.} reported \cite{Aarts2015} the observation of a colossal triplet spin-valve effect for the S/F1/N/F2 structure
made of amorphous MoGe, Ni, Cu, and CrO$_2$ as the S, F1, N, and F2 layers, respectively. This structure demonstrated variation of $T_c$ by $\sim$ 1
K when changing the relative alignment of the two F layers. It was shown that the optimal operational field for this device is of the order of 20
kOe. Gu {\it et al.} \cite{Gu1,Gu2} reported $\Delta T_c \sim 400$ mK for Ho/Nb/Ho trilayers. Also in this case the parallel configuration of
magnetizations was reached at a field of $\sim$ 10 kOe. The high operational fields of these spin valves are disadvantageous for the superconducting
spintronics. Besides, the physical reasons for large values of $\Delta T_c$ for spin valve based on half-metals are not yet theoretically explained.
This calls for elaboration of classical spin-valve structures which use standard ferromagnets (Fe, Co, Ni) and their alloys with good electrical
contacts between all layers and for which theoretical understanding of the operational principle is available.

Oh {\it et al.} \cite{Oh} proposed theoretically a metallic
F1/F2/S multilayer structure as a superconducting spin valve based
on the S/F proximity effect. In our previous works on the
CoO$_x$/Fe1/Cu/Fe2/In structure we have demonstrated a full
switching between the normal and superconducting states
\cite{leksin1} and observed the sign-changing oscillating behavior
of the magnitude of the spin valve effect $\Delta T_c$ on the
thickness of the Fe2 layer \cite{leksin2,leksin3}. With that the
F1/F2/S structure was experimentally established as a spin valve.

Recent theories (see, e.g., reviews \cite{bergeret,buzdin,efetov,efetov2,eschrig,blamire,Linder2015}) predict that at certain conditions  a
long-range triplet component (LRTC) in the superconducting condensate can arise in the S/F structure. The generation of the LRTC in the F1/F2/S
spin-valve structure should manifest itself  as a minimum of $T_c$ at noncollinear configuration of magnetizations \cite{Fominov2}. We have obtained
experimental confirmation of this prediction by studying the CoO$_x$/Fe1/Cu/Fe2/Pb spin valve structure \cite{leksin4}. The observed angular
dependence of $T_c$ was caused by a combination of the conventional and triplet components of the condensate.  An indication of the triplet
contribution to the magnitude of the superconducting spin-valve effect has been also observed  in Refs.
\cite{Zdravkov2013,Jara2014,Wang2014,Flokstra2015,Banerjee2014}.

A crucial question of fundamental and application-related
importance is whether it is possible to observe and even utilize
an ``isolated'' triplet spin-valve effect. At first glance, it
seems to be unrealistic, since LRTC arises entirely from the
singlet component and cannot exist without it. However, here we
experimentally demonstrate an ``isolation'' of the LRTC. We
achieved it by exploiting  the oscillatory behavior of $\Delta T_c
(d_{F2})$ that effectively suppresses the conventional spin-valve
effect by an appropriate choice of the F2-layer thickness
$d_{F2}$. Furthermore, we succeeded to utilize the LTRC for the
operation of the spin-valve and demonstrate the full switching
effect for the superconducting current upon changing the mutual
orientation of the magnetizations of F1 and F2 layers from AP to
the {\it orthogonal} orientation. Finally, we have substantially
improved the theoretical analysis by employing a fully
quantitative approach for calculation of the $T_c$ suppression due
to the proximity effect.

To investigate the proximity-induced triplet pairing we measured
the magnitude of the spin-valve effect in the dependence of the
Py2-layer thickness for the CoO$_x$/Py1/Cu/Py2/Cu/Pb multilayer
grown on the MgO (001) substrate. Here Py denotes permalloy
Ni$_{0.81}$Fe$_{0.19}$. The choice of Py for ferromagnetic layers
appears crucial. As will be shown below, Py due to a smaller value
of the exchange splitting of the conduction band in comparison
with Fe shifts the maximum of $\Delta T_c$ towards a larger
F-layer thickness and yields much larger $\Delta T_c$ for the
spin-singlet and spin-triplet channels. In addition, the use of Py
decreases the switching field by a factor of 6 in comparison with
Fe as a magnetic layer. This enables to avoid any possible partial
depinning of the bias Py1 layer by the switching field.

Optimization of the preparation conditions of the samples and
their characterization are described in Ref.~\cite{leksin_s}.
%
%
The optimal thickness of the Pb layer $d_{Pb}=70$ nm was
determined from the $T_c(d_{Pb})$ curve measured at a constant
$d_{Py1}=5$ nm, which is much larger than the penetration depth
$\xi_h$ of Cooper pairs into ferromagnetic Py. Basing on our data
on $T_c(d_{Py})$ at fixed $d_{Pb}$ we estimate this value as
$\xi_h \simeq 1.1$ nm. At a large Pb layer thickness, $T_c$ slowly
decreases with decreasing $d_{Pb}$. Below $d_{Pb} \sim 120$ nm,
$T_c$ value starts to decrease rapidly. At $d_{Pb} \leq 40$ nm,
$T_c$ is less than 1.5 K. At small $d_{Pb} \leq 70$ nm the width
of the superconducting transition curve gets extremely large, of
the order of 0.4 K. Bearing in mind that the influence of the
magnetic part of the structure gets stronger as the S-layer
thickness approaches the superconducting coherence length $\xi_s$
($\xi_s\simeq 40$ nm for our samples), we have chosen  $d_{Pb}=70$
nm as a compromise value.

Earlier we revealed that the F1-layer thickness at a fixed
$d_{F2}$ does not significantly influence $\Delta T_c$ for the
sample set
CoO$_x$ (2.5) / Fe1 ($d_{Fe1}$) / Cu (4) / Fe2 ($d_{Fe2}$) / Cu (1.2) / Pb (60 nm)
and that a thin Cu (1.2-nm) interlayer between F2 and S layers is
completely transparent for the Cooper pairs \cite{leksin5}.

All spin-valve structures were magnetically characterized using a standard 7-T VSM SQUID magnetometer (Fig.~1). First, the samples were cooled from 300 to 10\,K in the presence of the in-plane magnetic field $+4$ kOe. At 10 K the magnetic field was varied from $+4$ kOe to $-4$ kOe
and back again. During this variation, the in-plane magnetic moment of the sample was measured [Fig.~1(a)]. It turns out that for most of the
CoO$_x$/Py1/Cu/Py2/Cu/Pb samples the saturation field of the Py2 film is of the order of 200 Oe [Fig.~1(b)]. It can also be seen from this figure that
the switching field $H_0 = \pm 150$ Oe is sufficient to sustain a homogenous magnetization for the Py2 layer following the switching field direction
without formation of the domain structure \cite{leksin3}. At the same time the magnetization of the Py1 layer remains fixed up to the operating field
of the order of $-2.5$ kOe due to its pinning by the antiferromagnetic CoO$_x$ layer (N\'{e}el temperature $T_N\sim 250$ K). This result is very
similar to those observed for CoO$_x$/Fe1/Cu/Fe2/Cu/Pb structure (see Fig.~4 of Ref. \cite{leksin5}).

\begin{figure}[h]
\includegraphics[width=1.0\columnwidth,angle=-90]{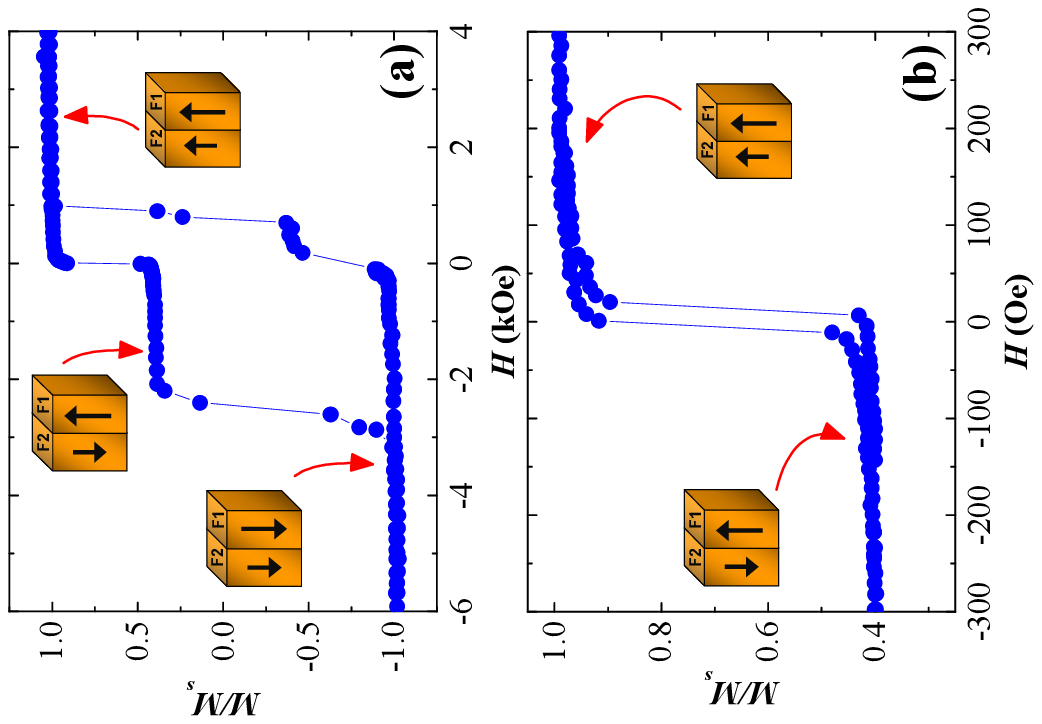}
\caption{(Color online). (a) Magnetic hysteresis loop for the CoO$_x$ (3) / Py (3 )/ Cu (4) / Py (1.2) / Cu (2) / Pb (70 nm) structure. (b)  The central part of the
minor hysteresis loop for this sample due to the reversal of the magnetization of the free Py2 layer. Arrows denote mutual orientation of the
magnetization of the two Py layers.}
\end{figure}

For the transport study we used another system which enables 
very accurate control of the magnetic field acting on the sample.
We have combined the electrical setup with a vector magnet that
enables a continuous rotation of the magnetic field in the plane
of the sample. To avoid the occurrence of the unwanted
out-of-plane component of the external field, the sample plane
position was always adjusted with an accuracy better than
$3^\circ$ relative to the direction of the dc external field. By
preserving the in-plane orientation of the external field we avoid
any noticeable angular dependent change in the demagnetization
field. The magnetic field value was measured with an accuracy of
$\pm 0.3$ Oe using a Hall probe.

For the set of spin-valve samples with various $d_{Py2}$, we studied
the dependence of the superconducting transition temperature $T_c$
on the angle $\alpha$ between the direction of the cooling field and
the external magnetic field both applied in plane of the sample. For
the CoO$_x$ (3) / Py (3) / Cu (4) / Py (0.6) / Cu(2) / Pb (70 nm) structure we
observed a large magnitude of  a conventional spin valve effect with
$\Delta T_c=110$ mK [Fig.~2(a)].

\begin{figure}[h]
\includegraphics[height=0.7\columnwidth,angle=-90]{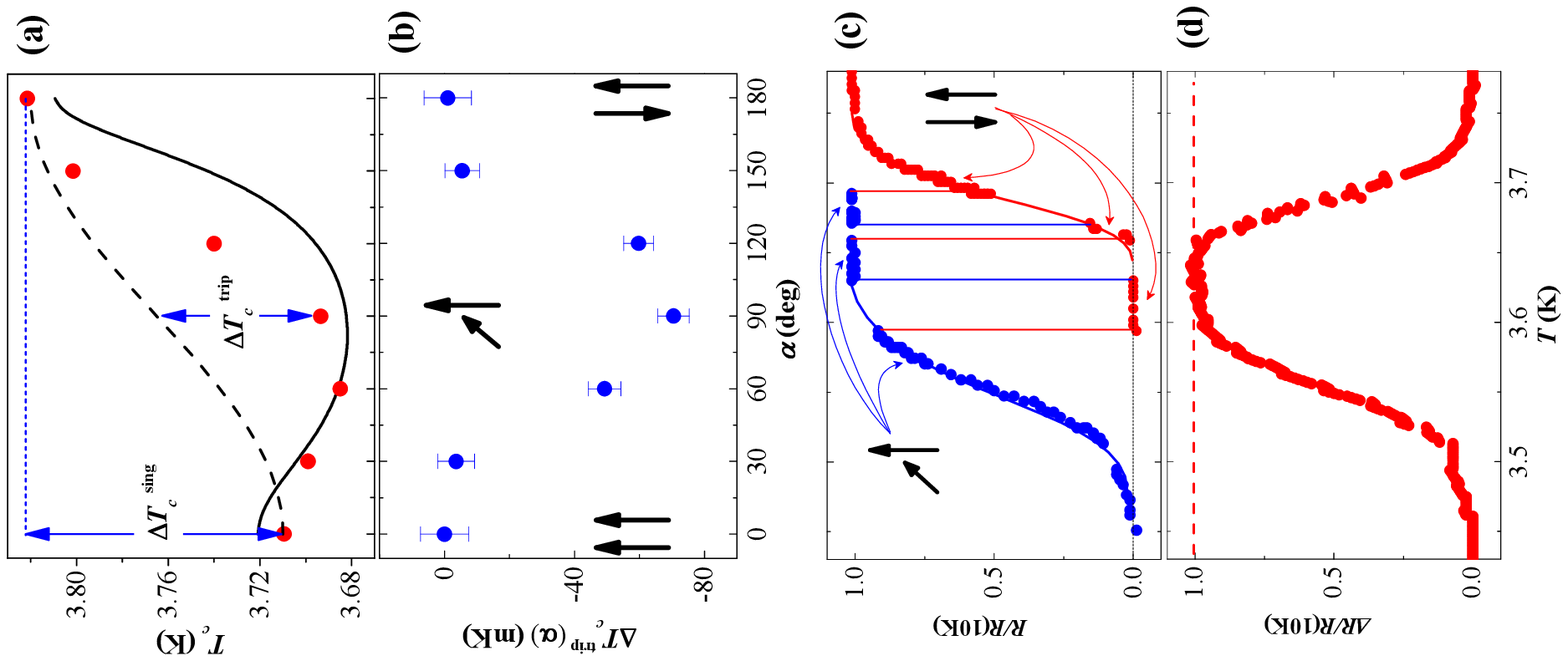}
\caption{(Color online).  Spin-valve effect for the
CoO$_x$ (3) / Py (3) / Cu (4) / Py(0.6) / Cu (2) / Pb (70 nm) structure. (a)
Angular dependence of $T_c$ measured at a field $H_0=150$ Oe
(circles), estimated $\alpha$ dependence of the singlet spin valve
effect $\Delta T_c^\mathrm{sing}$ (dashed line), and the theoretical
result according to Ref.~\cite{Fominov2} (solid line) (see the
text). (b) Difference $\Delta T_c^\mathrm{trip}$ between the
actual $T_c$ and the reference curve [dashed line in panel (a)]. (c)
Switching between normal and superconducting states in the
CoO$_x$ (2.5) / Py (3) / Cu (4) / Py (1.0) / Cu (2) / Pb (70) spin valve sample
by sweeping slowly the temperature within $\Delta T_c$ and
changing the direction of magnetic field $\alpha$ between
$180^\circ$ (closed circles) and $90^\circ$ (opened circles). (d)
Temperature dependence of $\Delta R=R(\alpha=90^
\circ)-R(\alpha=180^\circ)$ demonstrating the full switching in
the temperature range $3.6 \div 3.66$ K.}
\end{figure}

One can see in Fig.~2(a) that upon changing the mutual orientation
of magnetizations by a gradual in-plane rotation of the magnetic
field from the P ($\alpha =0^\circ$) to the AP ($\alpha
=180^\circ$) state, the $T_c$ does not change monotonically but
passes through a minimum. According to theory, a characteristic
minimum in the $T_c(\alpha)$ dependence (which is most pronounced
if it takes place near $\alpha = 90^\circ$ but may not be
necessarily located exactly at this angle) is a fingerprint of the
LRTC \cite{Fominov2}. Though the triplet component is inherent in
the case of noncollinear magnetizations, assuming for a moment its
absence, one would expect the $T_c(\alpha)$ dependence to be
monotonic. From general symmetry, $T_c(\alpha)$ must behave as
$\alpha^2$ and $(\pi -\alpha)^2$ when $\alpha$ deviates from $0$
and $\pi$, respectively (since deviations to both sides of this
values are physically equivalent, and we expect $T_c(\alpha)$ to
be an analytic function). One would then arrive at a simple
angle-dependent superposition of the limiting values of
$T_c^\mathrm{P}$ and $T_c^\mathrm{AP}$, which reads as
$T_c^\mathrm{(ref)}(\alpha)=T_c^\mathrm{P} \cos^2 (\alpha/2) +
T_c^\mathrm{AP} \sin^2 (\alpha/2)$. This dependence is shown by
the dashed line in Fig.~2(a) and we consider it as a reference
curve. Deviation of the actual $T_c$ from the reference curve,
shown in Fig.~2(b), demonstrates the LRTC contribution to the
$\Delta T_c$. We denote it as $\Delta T_c^\mathrm{trip}$. Figure~2(a)
shows that the difference in $T_c$ between AP and perpendicular
orientations of magnetizations amounts to as much as 130 mK. It
means that the LRTC significantly contributes to the spin-valve
effect.

In our previous works \cite{leksin2,leksin3,leksin4,leksin5}, we
compared our experimental data for $T_c$ with the effective
boundary parameter $W$ that enters the theory \cite{Fominov2} and
determines how strongly the F part of the system suppresses the
superconductivity in the S layer. This approach allowed us to
demonstrate a good qualitative agreement between theory ($W$) and
experiment ($T_c$) without actually calculating the critical
temperature. Here, we apply the method of Ref.~\cite{Fominov2}
(extended to the case of arbitrary S/F interface transparency
\cite{Deminov}) for direct fitting of our data on $T_c (\alpha)$,
as shown in Fig.~2(a) by a solid line. Here we used the following
set of parameters: the superconducting coherence lengths for the S
and the F layers $\xi_s =$ 41 nm and $\xi_f = $ 13 nm; the S layer
thickness $d_s = $ 73.5 nm; the bulk critical temperature of the S
layer $T_{cs}$ = 7.2 K; the transparency parameters $\gamma$ =
0.734 and $\gamma_b$ = 1.8; and the exchange field acting on the
spins of conduction electrons in the F layer $h=$ 0.3 eV. We found
that the theoretical description requires much smaller
$d_{Py2}=0.3$ nm than the nominal $d_{Py2}=0.6$ nm. There are
several possible reasons for that. First, the Py2 layer is
sandwiched by two Cu layers. Due to the interdiffusion the
effective thickness may be reduced down to 0.3 nm. Second, the
theory does not take into account details of the band structure of
the materials constituting our structure. It can also be that the
dirty-limit conditions assumed by the theory, are not fully
satisfied in our system. In any case, as seen in Fig.~2(a), theory
fits experimental data satisfactorily at reasonable values of
parameters.

The magnitude of the spin-valve effect upon changing the mutual
orientation of magnetizations from AP to the orthogonal one
exceeds the width of the superconducting transition curve.
Therefore it is possible to switch off and on the superconducting
current flowing through our samples {\it completely}, as
demonstrated in Fig.~2(c). A complete on/off switching of the
resistance of the sample due to the combination of the triplet
spin valve effect and ordinary spin valve effect is shown in
Fig.~2(d).

From the data in Fig.~2 it is obvious that the LRTC aids in
reaching larger spin-valve effect but still the triplet effect
interferes with the conventional one. As we have previously shown,
the amplitude of the conventional spin-valve effect can be
suppressed to zero for certain thickness $d_{F2}$ due to the
oscillating behavior of $\Delta T_c(d_{F2})$
\cite{leksin2,leksin3}. These oscillations are caused by the
interference in the F2 layer of the pair condensate wave function
coming from the F2/S interface with the one reflected from the
F1/F2 interface. For CoO$_x$/Fe1/Cu/Fe2/Cu/Pb structure  $\Delta
T_c$ reduces to zero for $d_{Fe2} \simeq 0.8 \div 1$ nm
\cite{leksin5,leksin6}.  From the analysis of our present data on
the $T_c(d_{Py})$ dependence we conclude that the penetration
depth of the Cooper pairs into Py exceeds the value obtained for
Fe by 30\%. Therefore $\Delta T_c=0$ in the samples with Py should
be reached at $d_{Py2} \sim 1 \div 1.7$ nm.  Indeed, such a sample
with $d_{Py}=1.7$ nm prepared in the present work demonstrates the
``isolated'' triplet spin-valve effect (Fig.~3).
\begin{figure}[h]
\includegraphics[height=0.7\columnwidth,angle=-90]{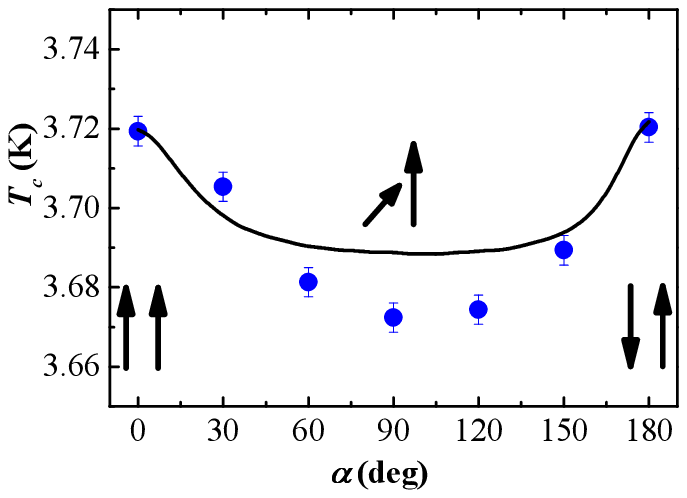}
\caption{(Color online).  Sample CoO$_x$ (3) / Py (3) / Cu (4) / Py (1.7) / Cu (2) / Pb (70) with the zero conventional spin-valve effect. Angular dependence of
$T_c$ caused by the LRTC is shown by circles, and the solid line is the theoretical curve according to Ref.~\cite{Fominov2} (see the text).}
\end{figure}
Obviously, the conventional singlet spin-valve effect vanishes,
$\Delta T_c=T_c^\mathrm{AP}-T_c^\mathrm{P} = 0$. At the same time
at noncollinear orientation of magnetizations of the Py1 and Py2
layers the $T_c(\alpha)$ dependence exhibits a minimum. In
accordance with theory \cite{Fominov2} for this sample the
amplitude of $\Delta T_c^\mathrm{trip}$ is smaller than for the
sample in Fig.~2 because of the larger value of $d_{Py2}$. That is
because the LRTC is generated when Cooper pairs reach the Py1
layer after leaving the superconductor and penetrating into the
Py2 layer. As a result, the thicker Py2 layer suppresses both the
singlet component and the LRTC that is generated from the singlet
one. Fitting of these data was performed using the parameters of
theory which we already used in calculation of the $\Delta T_c$
dependence in Fig.~2(a). The only difference was in $d_{Py2}$, which
was taken 1.7~nm.

The possibility to isolate the spin-triplet component which we have shown in this work is remarkable. The targeted engineering of S/F
heterostructures where peculiarities of the interference of the superconducting pairing wave function make the spin-singlet component ineffective
from the viewpoint of influencing $T_c$ appears to be straightforward. Such heterostructures are promising candidates for building spintronic devices
where the functionality of the triplet Cooper pairs carrying not only charge but also the spin polarization over a long distance is essential.
Generation, control, and manipulation of such spin supercurrents in S/F multilayers appear thus as an emerging field of undisputable importance
both for fundamental physics and for material research.

In conclusion, we have experimentally investigated the long-range
proximity-induced triplet superconductivity in the F1/F2/S
structure. Our main experimental result is the ``isolation'' of
the triplet spin-valve effect by exploiting the oscillating
behavior of the ordinary spin-valve effect. Furthermore, we have
shown that the spin-triplet component can be utilized for the
operation of the spin valve and demonstrate the spin-triplet-assisted full switching effect for the superconducting current. On
the theory side, we have successfully applied a fully quantitative
approach for calculating the suppression of $T_c$ due to the
proximity effect in the studied heterostructures.

%
%
%
%

This work was supported by the Deutsche Forschungsgemeinschaft
through Grant LE 3270/1-1. It  was also partially supported by
RFBR (Grants No. 13-02-01389-a and No. 14-02-00350-a), by programs of
the RAS, by the Ministry of Education and Science (Russian
Federation), and by the program ``5top100.''


\begin{thebibliography}{99}

\bibitem{eschrig}
M.\ Eschrig,
Phys.\ Today \textbf{64}(1), 43 
(2011).

\bibitem{blamire}
M.~G.\ Blamire and J.~W.~A.\ Robinson,
J.~Phys.: Condens.~Matter \textbf{26}, 453201 
(2014).

\bibitem{Linder2015}
J.\ Linder and J.~W.~A.\ Robinson, Nat. Phys. \textbf{11}, 307 (2015).

\bibitem{Zdravkov2013}
V.~I.\ Zdravkov, J.\ Kehrle, G.\ Obermeier, D.\ Lenk, H.-A.\ Krug von
Nidda, C.\  M\"{u}ller, M.~Yu.\ Kupriyanov, A.~S.\ Sidorenko,
S.\ Horn, R.\ Tidecks, and L.~R.\ Tagirov,
Phys.\ Rev.~B \textbf{87}, 144507 
(2013).

\bibitem{Jara2014}
A.~A.\ Jara, C.\ Safranski, I.~N.\ Krivorotov, C.-T.\ Wu, A.~N.\
Malmi-Kakkada, O.~T.\ Valls, and K.\ Halterman,
Phys.\ Rev.~B \textbf{89}, 184502 (2014).

\bibitem{Wang2014}
X.~L.\ Wang, A.\ Di~Bernardo, N.\ Banerjee, A.\ Wells,
F.~S.\ Bergeret, M.~G.\ Blamire, and J.~W.~A.\ Robinson,
Phys.\ Rev.~B \textbf{89}, 140508(R) 
(2014).

\bibitem{leksin1}
P.~V.\ Leksin, N.~N.\ Garif'yanov, I.~A.\ Garifullin,  J.\ Schumann,
H.\ Vinzelberg, V.\ Kataev, R.\ Klingeler,  O.~G.\ Schmidt, and
B.\ B\"{u}chner,
Appl.\ Phys.\ Lett.\ \textbf{97}, 102505 (2010).

\bibitem{Li} Bin Li, N.~Roschewsky, B. A. Assaf, Marius Eich, M.~Epstein-Martin, D.~Heiman, M.~Munzenberg,
and J. S.~Moodera, Phys.~Rev.~Lett. {\bf 110}, 097001 (2013).

\bibitem{Aarts2015}
A.\ Singh, S.\ Voltan, K.\ Lahabi, and J.\ Aarts, Phys.\ Rev.~X \textbf{5}, 021019 (2015).

\bibitem{Gu1}
Y.\ Gu, J.~W.~A.\ Robinson, M.\ Bianchetti, N.~A.\ Stelmashenko, D.\ Astill,
F.~M.\ Grosche, J.~L.\ MacManus-Discoll, and M.~G.\ Blamire, APL Mat.\ \textbf{2}, 046103 (2014).

\bibitem{Gu2}
Y.\ Gu, G.~B.\ Hal\'{a}sz, J.~W.~A.\ Robinson, and M.~G.\ Blamire,
Phys.\ Rev.\ Lett.\ \textbf{115}, 067201 (2015).

\bibitem{Oh}
S.\ Oh, D.\ Youm, and M.~R.\ Beasley, Appl.\ Phys.\ Lett.\ \textbf{71}, 2376 (1997).



\bibitem{leksin2}
P.~V.\ Leksin, N.~N.\ Garif'yanov, I.~A.\ Garifullin, J.\ Schumann,
V.\ Kataev,  O.~G.\ Schmidt, and B.\ B\"{u}chner,
Phys.\ Rev.\ Lett.\ \textbf{106}, 067005 (2011).

\bibitem{leksin3}
P.~V.\ Leksin, N.~N.\ Garif'yanov, I.~A.\ Garifullin, J.\ Schumann,
V.\ Kataev, O.~G.\ Schmidt, and B.\ B\"{u}chner,
Phys.\ Rev.~B \textbf{85}, 024502 
(2012).

\bibitem{bergeret}
F.~S.\ Bergeret, A.~F.\ Volkov, and K.~B.\ Efetov,
Rev.\ Mod.\ Phys.\ \textbf{77}, 1321 
(2005).

\bibitem{buzdin}
A.~I.\ Buzdin, Rev.\ Mod.\ Phys.\ {\bf 77}, 935 
(2005).

\bibitem{efetov}
K.~B.\ Efetov, I.~A.\ Garifullin, A.~F.\ Volkov, and
K.\ Westerholt,
{\it Proximity Effects in Ferromagnet/Superconductor Heterostructures: Magnetic
Heterostructures: Advances and Perspectives in Spinstructures and
Spintransport, Series Springer Tracts in Modern Physics, Vol. 227 (Berlin, Springer, 2007)}, pp. 251-289.

\bibitem{efetov2}
K.~B.\ Efetov, I.~A.\  Garifullin, A.~F.\ Volkov, and
K.\ Westerholt, {\it Spin-Polarized Electrons in the
Superconductor/Ferromagnet Hybrid Structures: Magnetic
Nanostructures: Spin Dynamic and Spin Transport, Series Springer
Tracts in Modern Physics, Vol. 246, edited by H. Zabel and
M. Farle (Springer-Verlag, Berlin, 2013)}, pp. 85-118.
\bibitem{Fominov2}
Ya.~V.\ Fominov, A.~A.\ Golubov, T.~Yu.\ Karminskaya,
M.~Yu.\ Kupriyanov, R.~G.\ Deminov, and L.~R.\ Tagirov,
Pis'ma Zh.\ Eksp.\ Teor.\ Fiz.\ \textbf{91}, 329 
(2010) [JETP Lett.\ \textbf{91}, 308 
(2010)].

\bibitem{leksin4}
P.~V.\ Leksin, N.~N.\  Garif'yanov, I.~A.\ Garifullin,
Ya.~V.\ Fominov, J.\ Schumann, Y.\ Krupskaya, V.\ Kataev,
O.~G.\ Schmidt, and B.\ B\"{u}chner,
Phys.\ Rev.\ Lett.\ \textbf{109}, 057005 
(2012).

\bibitem{Flokstra2015}
M.~G.\ Flokstra, T.~C.\ Cunningham, J.\ Kim, N.\ Satchell, G.\ Burnell, P.~J.\ Curran,
S.~J.\ Bending, C.~J.\ Kinane, J.~F.~K.\ Cooper, S.\ Langridge, A.\ Isidori, N.\ Pugach,
M.\ Eschrig, and S.~L.\ Lee,
Phys.\ Rev.~B \textbf{91}, 060501 (2015).

\bibitem{Banerjee2014}
N.\ Banerjee, C.~B.\ Smiet, R.~G.~J.\ Smits, A.\ Ozaeta, F.~S.\
Bergeret, M.~G.\ Blamire, and J.~W.~A.\ Robinson,
Nature Communications\ \textbf{5}, 3048 (2014).

\bibitem{leksin_s}
P.~V.\ Leksin, A.~A.\ Kamashev, J.\ Schumann, V.~\ Kataev, J.\
Thomas, B.\ B\"{u}chner, I.~A.\ Garifullin, Nano Res., arXiv:1510.04846 (accepted).
\bibitem{leksin5}
P.~V.\ Leksin, N.~N.\ Garif'yanov, A.~A.\ Kamashev, Ya.~V.\ Fominov, J.\ Schumann,
C.\ Hess, V.\ Kataev, B.\ B\"{u}chner, and I.~A.\ Garifullin,
Phys.\ Rev.~B \textbf{91}, 214508 (2015).

\bibitem{Deminov}
R.~G.\ Deminov, L.~R.\ Tagirov, R.~R.\ Gaifullin, T.~Yu.\ Karminskaya,
M.~Yu.\ Kupriyanov, Ya.~V.\ Fominov, and A.~A.\ Golubov,
J.~Magn.\ Magn.\ Mater.\ \textbf{373}, 16 (2015);
R.~G.\ Deminov, L.~R.\ Tagirov, R.~R.\ Gaifullin, Ya.~V.\ Fominov, T.~Yu.\ Karminskaya,
M.~Yu.\ Kupriyanov, and A.~A.\ Golubov, Solid State Phenom. \textbf{233-234}, 745 (2015).


\bibitem{leksin6}
P.~V.\ Leksin, A.~A.\ Kamashev, N.~N.\ Garif'yanov,
I.~A.\ Garifullin, Ya.~V.\ Fominov, J.\ Schumann, C.\ Hess, V.\ Kataev, and B.\ B\"{u}chner,
Pis'ma Zh.\ Eksp.\ Teor.\ Fiz.\ {\bf 97}, 549 
(2013) [JETP Lett.\ \textbf{97}, 478 
(2013)].

\end{thebibliography}
\end{document}